# Monolayer Capping Provides Close to Optimal Resistance to Laser Dewetting of Au Films

Christopher P. Murray,* Daniyar Mamyraimov, Mugahid Ali, Clive Downing, Ian M. Povey, David McCloskey, David D. O'Regan, and John F. Donegan

ACCESS | Metrics & More | Article Recommendations | Supporting Information

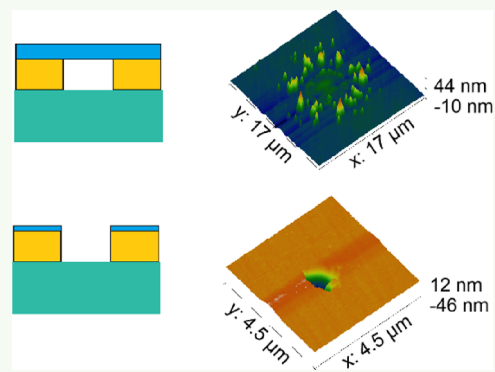

**ABSTRACT:** Next-generation heat-assisted magnetic recording (HAMR) relies on fast, localized heating of the magnetic medium during the write process. Au plasmonic near-field transducers are an attractive solution to this challenge, but increased thermal stability of Au films is required to improve long-term reliability. This work compares the effect of nanoscale Al, $AlO_x$, and Ta capping films on Au thin films with Ti or Ta adhesion layers for use in HAMR and other high-temperature plasmonic applications. Thermal stability is investigated using a bespoke laser dewetting system, and SEM and AFM are extensively used to interrogate the resulting dewet areas. The most effective capping layers are found to be 0.5−1 nm of Al or $AlO_x$, which can eliminate dewetting under certain conditions. Even one monolayer of $AlO_x$ is shown to be highly effective in reducing dewetting. In the case of thicker capping layers of Ta and $AlO_x$, the Au film can easily dewet underneath, leaving an intact capping layer. It is concluded that thinner capping layers are most effective against dewetting as the Au cannot dewet without breaking them and pulling them apart during the dewetting process. A simple model based on energetics considerations is developed, which explains how thinner capping layers can more effectively protect the metal from pore or fissure creation. The model provides some convenient guidelines for choosing both the substrate and capping layer, for a given metal, to maximize the resistance to laser-induced damage.

## ■ INTRODUCTION

Heat-assisted magnetic recording (HAMR) is a commercially important technology predicted to increase hard drive storage capacities to 50 TB and beyond in the coming years. Before entering the market, a range of technical and material challenges must be overcome, one of which is a reliable means of locally heating the high-coercivity perpendicular recording medium to c. 450 °C to write data.[1] To achieve this, plasmonic near-field transducers, which use Au thin films to couple laser energy to the medium, are favored.[2] Due to thermal stress over time, the Au films can dewet from their substrate even when adhesion layers are employed, leading to device failure. Reducing the likelihood of dewetting under normal operating conditions would therefore add to long-term device reliability.

Solid-state dewetting is a process by which a thin film in a metastable state relaxes and agglomerates into a lower energy configuration when supplied with thermal energy.[3] The process can nucleate at grain boundaries, film/air or film/substrate interfaces for example,[4,5] or their triple junctions. As dewetting happens in the solid state, it can occur at temperatures well below the melting point of the material in question. While some applications take advantage of the dewetting process to form nanoscale features[6−11] (catalysis,[12] magnetic recording,[13] sensors,[14] etc.), it becomes problematic when continuous thin films are required. Indeed, this has been an issue in the development of modern microelectronics where nanoscale films are used in applications where thermal budget must be considered.[15] It is particularly true in the case of Au thin films, which have considerable utility in plasmonics, but which adhere poorly to oxide substrates due to chemical inertness. Adhesion layers are commonly deployed to counter the likelihood of dewetting of Au from a substrate, although choosing the optimal material and thickness of adhesion layer is important.[16,17] In a previous study, we showed that ultrathin adhesion layers of about 0.5 nm thickness provided the best resistance to dewetting for Au films of 50 nm thickness.[18] Adhesion layers of 5 nm, which are commonly used in many processes, were found to dewet much more quickly than those with 0.5 nm adhesion layers. Similarly, applying capping layers to thin film surfaces can reduce dewetting by modifying the surface energy and, potentially more significantly, the ionic diffusion coefficients. Capping layers have been shown to help prevent polymer thin films from dewetting, and the same is true for Au nanostructures.[19−21] As with adhesion layers, when $AlO_x$ was applied as a capping layer, the performance of thinner layers is superior to thicker layers in preventing dewetting.[19] We have recently shown that depositing nanoscale Al capping layers on the surface of Ta 0.5 nm/Au 50 nm films has a positive impact on its resistance to dewetting.[22] It was found that thinner (0.5 nm) Al capping layers, which oxidize in air to $AlO_x$, conferred superior dewetting resistance compared to thicker variants (up to 5 nm). The reason for this enhanced dewetting resistance for ultrathin capping layers was not well understood.

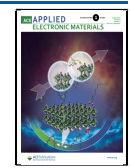

In a particularly interesting example, Cao et al. transferred monolayer graphene to cap Au films and found a dramatic reduction in the propensity of the film to dewet.[23] It was proposed that the enormous stiffness of graphene, which has a Young's modulus of 1 TPa, is the key contributor to this result. Simple mathematical models were developed, which simulated the dewetting of the films, calculating the surface energies of each surface and the strain in the capping layer through the process. Thus, the total energy of the system could be found, allowing the lowest energy end-state geometry to be predicted. Unfortunately, graphene transfer is laborious and remains unlikely to become an industrially viable process in the near term. Ideally, capping layers should be developed, which can be deposited with precise thickness control by common industrial processes such as sputtering. This is the focus of our work.

Initially, the effects of a sputter deposited capping layer material M = Al, $AlO_x$, or Ta on the dewetting characteristics of Ta 0.5 nm/Au 50 nm films are investigated. These capping materials were chosen as their bulk melting points and Young's moduli cover a large range[24] as shown in Table 1. We can vary the capping thickness in a way not easily available for the graphene study.

Table 1. Melting Point and Young's Moduli of Bulk Materials

| material | melting point (K) | Young's modulus (GPa) |
| --- | --- | --- |
| Al | 933 | 70 |
| $AlO_x$ | 2288 | 345−409 |
| Au | 1337 | 79 |
| Ta | 3290 | 186 |
| $Ta_2O_5$ | 2145 | 140 |

A schematic of the final film configuration is shown in Figure 1. The target thickness of the sputter-deposited capping layers is 0.5 nm ≤ $t$ ≤ 5 nm. These thicknesses were not measured directly but were based on timed depositions calculated from thicker measurable films made using the same conditions. It is unlikely that complete surface coverage is achieved for thin capping layers, particularly since the samples were then stored in air upon removal from the sputtering system. It is therefore most likely that capping layers of 1 nm and less are discontinuous and have more island-like morphology compared to thicker variants. Metallic Al oxidizes readily under ambient conditions and generally forms a limiting, amorphous, near-stoichiometric $Al_2O_3$ overgrowth up to a thickness of ≈4 nm.[25,26] This means that Al capping layers where $t ≤ 2$ nm are expected to be fully oxidized, while the 5 nm variant is not. Indeed, this difference is clearly seen with the naked eye as remaining metallic Al in the 5 nm capping layer is highly reflective at visible wavelengths. The $AlO_x$-capped samples are deposited from $Al_2O_3$ ceramic target-facing-targets using RF sputtering. They are assumed to be near-stochiometric $Al_2O_3$. Ta oxidizes readily in air to form mixed suboxides of $Ta_2O_5$ for all film thicknesses, and indeed, there is no visible difference between the samples.[27] For HAMR applications, the presence of an electrically insulating oxide capping layer is not an impediment as power is coupled to the layer optically.

The UV/vis transmission and reflectance spectra were recorded for each sample, and the absorption of each film was calculated. The dewetting characteristics of the samples were measured by collecting the back-reflected signal of a 488 nm laser heat-source, using a customized microscope experimental setup previously described.[28] By adjusting the output power of the laser to match the absorption of the film, the dewetting response was measured 3 times for each sample at absorbed powers $P_{abs}$ = 30, 35, and 40 mW, typically for a period of 300 s. The $1/e^2$ beam waist was also recorded for each measurement and was on average $\mu$ = 1.18 $\mu$m with a standard deviation $\sigma$ = 0.06 $\mu$m. The resulting temperature rise in the films is about 375−500 K over this $P_{abs}$ range.[29] SEM and AFM analyses were carried out to characterize the subsequent dewetting response of the films after dewetting.

## RESULTS/DISCUSSION

Figure 2a shows the absorption of the samples as measured using UV/vis spectrometry at $\lambda$ = 488 nm. The absorption is roughly similar for Al 0.5 nm ≤ $t$ ≤ 2 nm but decreases for the $t$ = 5 nm case. This is due to increased reflectance of the Al 5 nm capped sample, which is not fully oxidized. The increase in reflectance causes the absorption to reduce. The Ta-capped sample absorption behaves rather linearly with thickness, but not so for $AlO_x$, which shows a minimum at a thickness of 2 nm.

The numerical calculations of the absorption are studied using the finite element method (FEM) (COMSOL Multiphysics software in 2D mode). The computation region consists of a gold film with a uniform thickness of 50 nm on top of a glass substrate of a semi-infinite thickness (Figure 2b). The capping layer thickness varied from 0.5 to 5 nm. The model includes periodic boundary conditions on the left and right sides of the computation space to simulate an infinite film width and port boundary conditions at the top and bottom. The port boundary condition (port 1) at the top launches a laser power of 1 W/m at $\lambda$ = 488 nm perpendicular to the surfaces of the film stack. Port 2 is the exit port. We assumed in this model that all the surfaces are perfectly flat, such that it exhibits negligible structural alterations in the plane of the gold surface. Therefore, it can be modeled quite easily by considering a small 2D unit cell with an arbitrary width much smaller than the wavelength. Nevertheless, as metals have wavelength-dependent refractive indices, the mesh size is adjusted manually according to the minimum wavelength in each material as well as the skin depth. The discrepancy of the simulations versus the experiment graph for Ta was a result of the oxidation of the Ta thin film (Figure 2c). A trend is obtained consistent with experimental data when we replaced the Ta with $Ta_2O_5$ material in simulations.

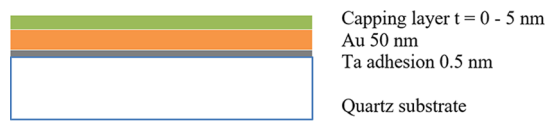

Figure 1. Schematic of film stack under examination.

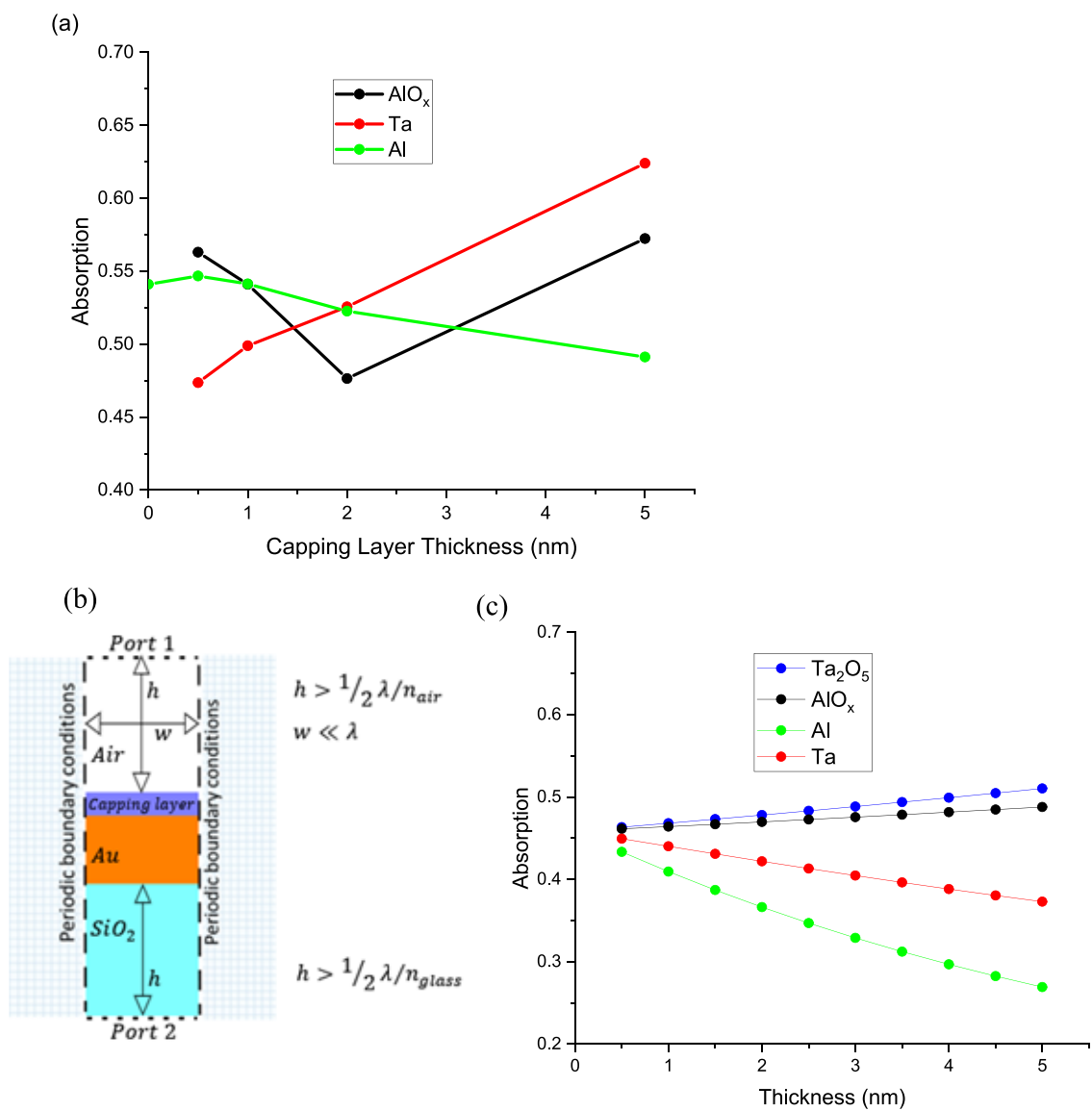

**Figure 2.** (a) Experimentally measured sample power absorption vs capping layer type and thickness, (b) numerical COMSOL model used to predict (c) absorption as a fraction of incident power at $\lambda$ = 488 nm vs capping layer type and thickness.

In Figure 3a, normalized dewetting reflectance curves for a Ta 0.5 nm/Au 50 nm film with no capping layer are shown for $P_{abs}$ = 30, 35, and 40 mW. As the film dewets under the laser, the back-reflected signal decreases over time at a rate, which depends on the power absorbed by the sample. The greater the power absorbed, the faster and greater the dewetting and consequently the greater the total reduction in reflectance. The reflectance change $\Delta R$ is defined as the percentage change in reflectance with respect to its initial value after a defined period of heating of 300 s. SEM was then used to image the dewet areas. The dewet area was estimated from SEM images using ImageJ analysis software. A trend can clearly be observed in the dewet area—increased power absorbed by the film results in larger dewet areas (Figure 3b). The results, the inset in the images, are plotted against $\Delta R$ in Figure 3c. The correlation is not perfect ($R^2$ value of 0.93) as the edges of the dewet areas are very rough. This is indicative of the large Au grain edges, which have developed because of localized heating. One can reasonably conclude however that the change in reflectance $\Delta R$ is a good indicator of the size of the dewet area on the sample. We use both to characterize our samples. A video recording of the dewetting process is shown in the supplementary materials where the development of the dewetting process under laser irradiation can be seen.

Similar reflectance change curves were also recorded for each capped sample. Three measurements were made for each combination of capping thickness and laser power. The results are summarized in Figure 4 as a set of response surfaces. In these diagrams, the greater the change in reflectance, the greater the trend toward red. It is evident that, as expected, $\Delta R$ is greater at higher $P_{abs}$ for all capping layers. Al-capped samples showed the largest range of responses. The least dewetting, with $\Delta R$ = 0%, was found when $t$ = 0.5 nm, $P_{abs}$ =

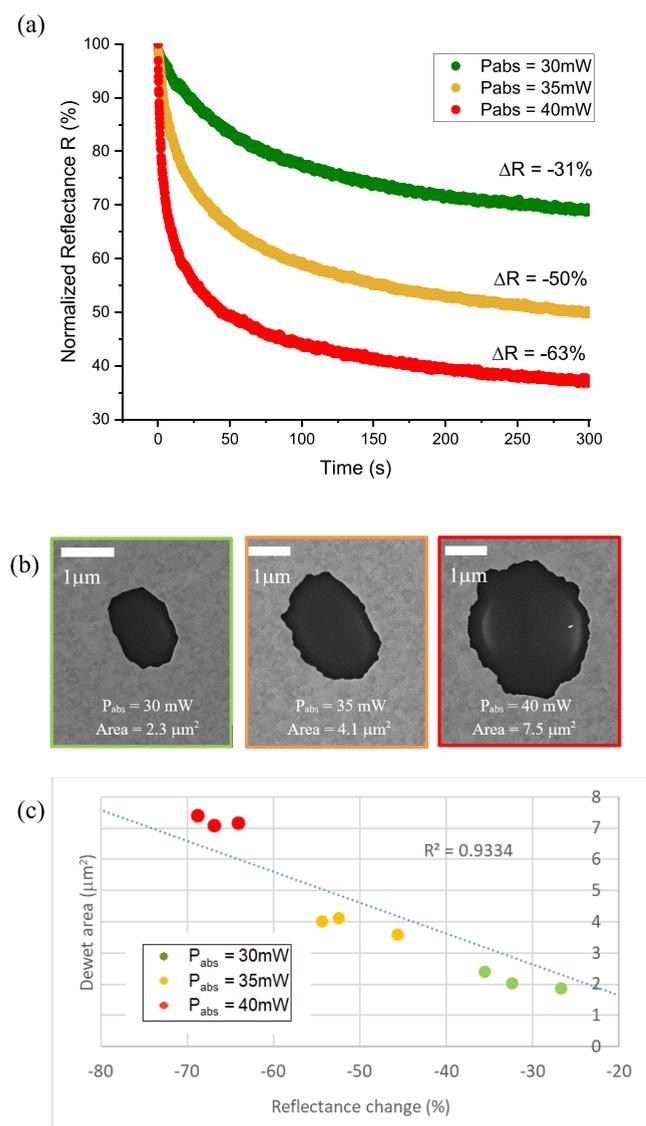

Figure 3. (a) Normalized dewetting curves obtained for uncapped Ta 0.5 nm/Au 50 nm sample at various absorbed powers $P_{abs}$, (b) the resulting damage to the films with the damage areas inset, and (c) the relationship between the change on reflectance and the resulting dewet area.

30 and 35 mW. The worst dewetting, $\Delta R = 66-77\%$, occurred when $t = 5$ nm. The $AlO_x$-capped samples were generally not quite as robust as Al-capped samples, but in some circumstances were marginally better. Ta-capped samples performed worst, with $\Delta R$ changes of up to 90%. Most surprisingly, thinner capping layers performed better than thicker variants for all capping materials, with capping layers of 0.5−1 nm offering the greatest protection against dewetting in all cases. In some cases, dewetting was prevented entirely. We found similar behavior for adhesion layers in a previous study in that the 0.5 nm adhesion also produced the maximum in dewetting resistance[11] but the explanation for this behavior cannot be the same.

To further understand this behavior, SEM was used to image the dewet areas of capped samples, and some typical examples are shown in Figure 5 for $P_{abs} = 40$ mW. It is again clear that the thicker the capping layer, the greater the areas dewet. For 5 nm thick capping layers, hillock formation is observed in the case of Ta and $AlO_x$, but the Al capping layer appears to blister more at the edge of the dewet area. This may be related to the fact that there remains metallic Al in this layer, whereas the other capping materials are more likely to be fully composed of oxides. As the capping layer becomes thinner, the dewet area reduces. The general shape of the dewet area becomes rougher. This roughness is likely following the Au grain edges, which have grown due to heating.

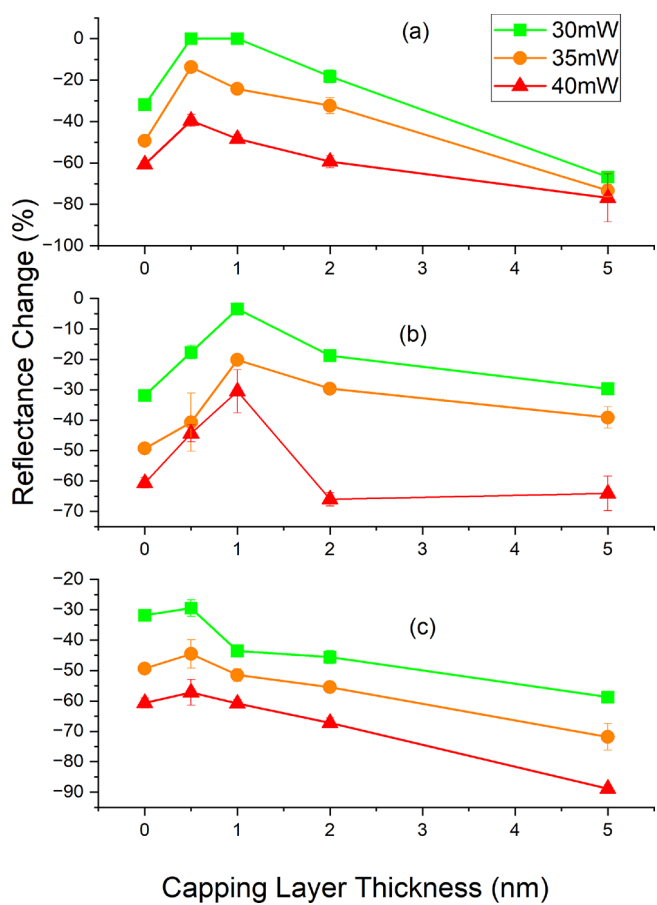

Figure 4. Response of reflectance change $\Delta R$ vs target capping layer thickness $t$ and power absorbed $P_{abs}$ for capping layers sputtered from (a) Al, (b) $AlO_x$, and (c) Ta targets. Error bars represent standard error.

For each SEM image, the dewet area was calculated using image analysis in ImageJ software. The results are shown in Figure 6 for each capping layer vs $P_{abs}$. In some cases, where no $\Delta R$ was recorded, there is no observable dewet area under SEM and the area value is therefore zero.

Again, thinner capping layers offer superior dewetting resistance than thicker capping layers, irrespective of material choice. On average, the Al = 0.5 nm capped samples offer the best protection against dewetting across the $P_{abs}$ range tested, with no observable dewetting at $P_{abs} = 30$ and 35 mW. This agrees with data presented in Figure 4. In contrast, Al = 5 nm capping was the worst performer overall. The range of dewet areas is tighter for Ta and $AlO_x$ with $AlO_x = 1$ nm resisting dewetting at $P_{abs} = 30$ mW. Once oxidized, the Al = 0.5 nm thick layer becomes $AlO_x \approx 1$ nm, so this result is consistent. The overall protective effect of thin capping layers is illustrated in Figure 7, where the best performing capped sample is contrasted with an uncapped sampled. Dewetting is profoundly reduced for the capped sample.

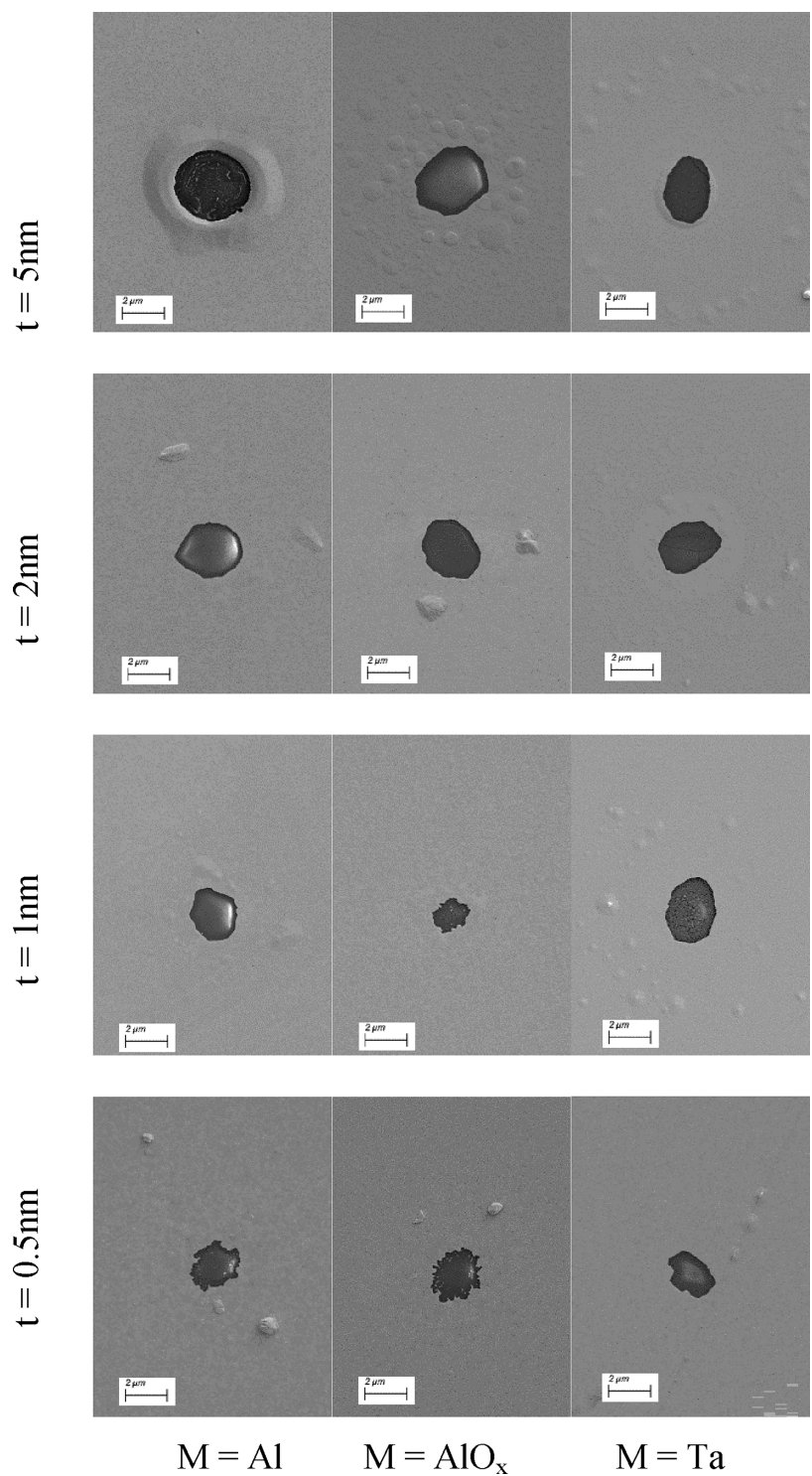

**Figure 5.** Typical SEM images of capped samples post dewetting at $P_{abs}$ = 40 mW for capping materials M = Al, AlO$_x$, and Ta (horizontal) and capping thicknesses 0.5−5 nm (vertical).

Close inspection of the SEM images offers some insight into the possible reasons for this behavior. For some Ta capping layers of 5 and 2 nm, cracks across the dewet area are observed, suggesting some material remains (Figure 8). SEM/EDX and AFM were used to examine these samples. At low accelerating voltage (2 kV) and by tilting the sample, one can image what looks like a membrane of remaining capping layer covering the dewet area (Figure 9a). Elemental mapping using EDX was then used to scan this area for the characteristic X-ray emission peaks of Au (M$_\alpha$ peak at 2.12 keV) and Si (K$_\alpha$ peak at 1.739 keV). The absence of Au is confirmed in this area, associated with a stronger Si signal coming from the substrate (Figure 9b). This means that Au dewetting has occurred. However, AFM mapping of the area reveals a membrane remains intact across the dewet area, with some hillocks around the edges (Figure 9c). Finally, a line section through this area is shown in Figure 9d confirming the membrane is somewhat deformed but is intact. Further AFM measurements on selected samples indicate the membrane remains intact, but occasionally sagging, on all but the Ta = 0.5 nm sample (Table 2). In this instance, the AFM probe penetrated almost to the substrate surface. While hillock formation may be caused by compressive stresses in the film resulting from thermal expansion differences during annealing,[30] we have previously shown that the volume of the hillocks in fact is almost equal to the volume of dewet material.[22] This shows that diffusion of Au to the hillocks is taking place and that evaporation or ablation plays no role here.

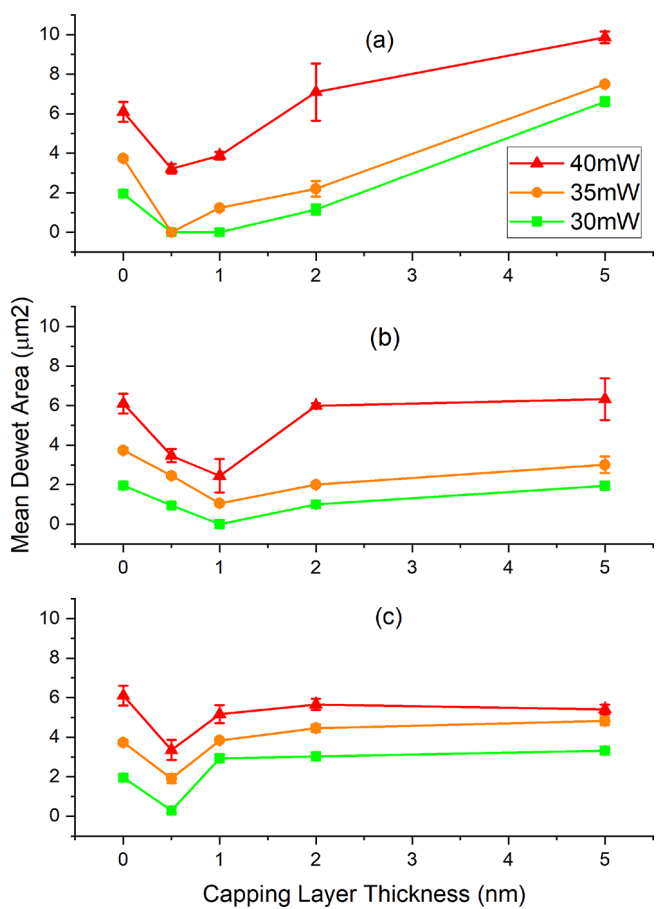

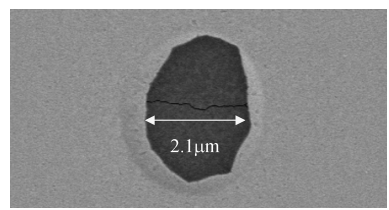

**Figure 8.** Ta (5 nm)-capped sample post laser dewetting at $P_{abs}$ = 40 mW. Note what appears to be a crack across the dewet area.

**Figure 6.** Mean dewet area vs capping layer thickness for each $P_{abs}$ and capping layer: (a) Al, (b) AlO$_x$, and (c) Ta. Error bars represent standard error.

Further AFM analysis of selected samples indicates that this type of dewetting under 5 nm capping layer membranes can occur also for AlO$_x$ (Figure S1), but not for Al where unique behavior is observed (Figure S2). No surviving membranes were discovered on any sample where the capping thickness was $t$ = 0.5 nm (Figure S3). This suggests a correlation between gold layer dewetting between capping layer and substrate, and the dewetting resistance of the stack.

To further investigate these effects with finer control of capping layers thickness, atomic layer deposition (ALD) was used for capping layer deposition. ALD allows for fine control of deposition. Using trimethyl aluminum and water as precursors, each set of pulses typically delivers 1.1−1.2 Å of Al$_2$O$_3$ onto the sample, which is less than one monolayer.[31] A new series of quartz/Ti 1 nm/Au 42 nm samples were prepared by electron beam evaporation, which were then capped with up to 20 monolayers of Al$_2$O$_3$. Again, these samples were laser dewet at $P_{abs}$ = 30 mW and examined as above.

Figure 10 shows a plot of normalized reflectance change $\Delta R$ vs number of ALD Al$_2$O$_3$ pulses $N$ with $P_{abs}$ = 30 mW. Three measurements per sample were made. The largest $\Delta R$ occurs for the uncapped sample, as expected. However, the application of just one ALD pulse of Al$_2$O$_3$ has reduced $\Delta R$ significantly, indicating less dewetting has occurred. Additional layers of Al$_2$O$_3$ add further protection against dewetting, with $N$ = 5, or a capping layer thickness of c. 0.5 nm, offering optimal protection before tailing off slightly at $N$ = 20.

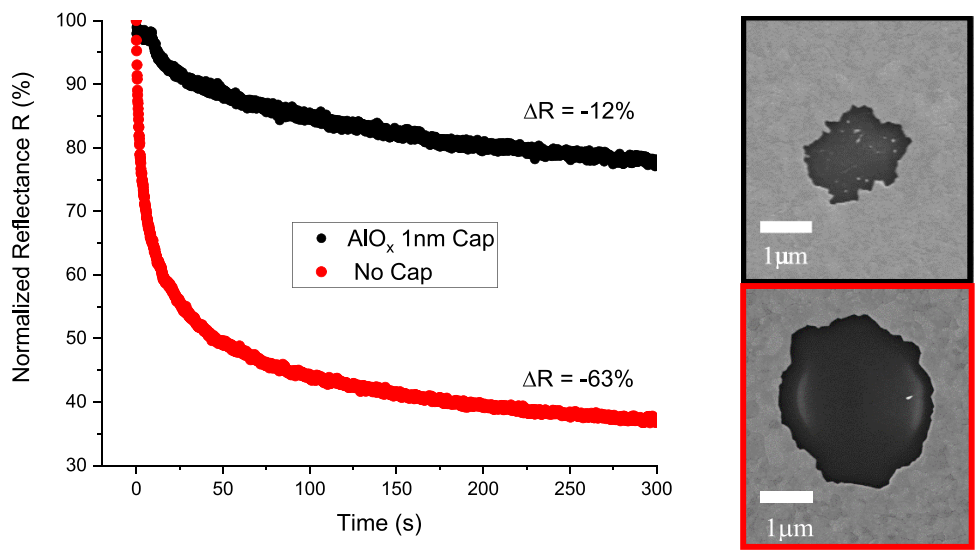

**Figure 7.** Best capped sample vs uncapped sample dewetting reflectance curves at $P_{abs}$ = 40 mW and (inset) resulting dewet area SEM images. The capped sample offers superior resistance to dewetting.

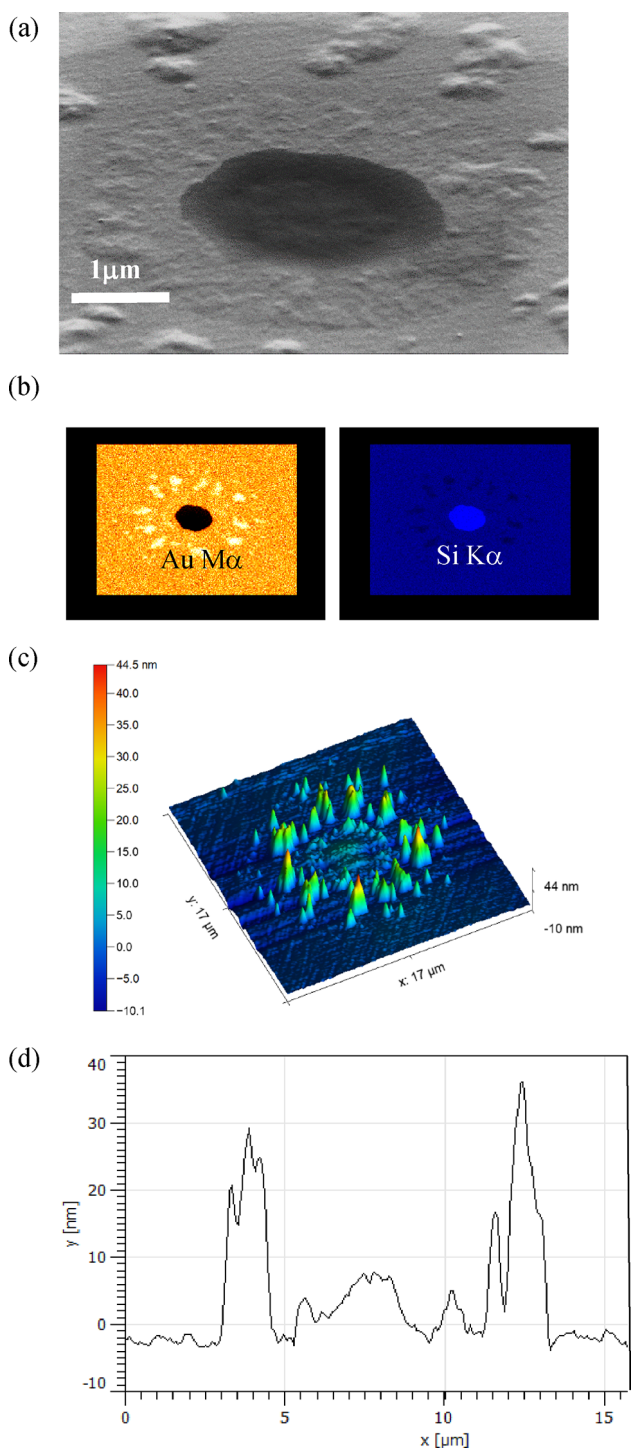

**Figure 9.** (a) Low accelerating voltage (2 kV)-tilted SEM image of Ta 0.5 nm, Au 50 nm, and Ta 5 nm capped sample post dewetting at $P_{abs}$ = 30 mW, (b) EDX elemental maps of the dewet area, the image on the left shows Au is missing from the center, while image on the right shows Si is present as would be expected for the substrate, (c) AFM topographical image of the dewet area, and (d) a line section across the center of this image showing membrane is intact.

**Table 2. Maximum AFM Tip Deflection on the Ta Membrane Post Dewetting, Measured Relative to the Sample Surface**

| Ta cap thickness (nm) | AFM deflection (nm) |
|---|---|
| 0.5 | −45 |
| 1 | −9 |
| 2 | −6 |
| 5 | +10 |

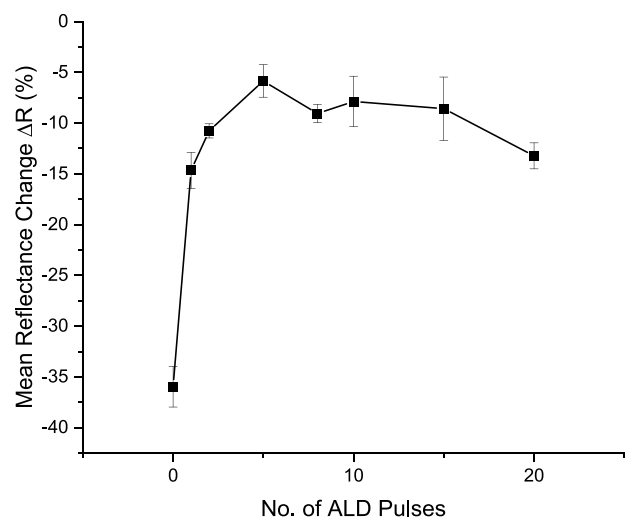

**Figure 10.** Plot of reflectivity change $\Delta R$ vs number of ALD pulses $N$ for Ti 1 nm/Au 42 nm/Al$_2$O$_3$ samples with $P_{abs}$ = 30 mW. Error bars represent standard error.

Figure 11a shows SEM imagery of the dewet areas, confirming that the maximum dewetting has occurred where no capping was applied. Dewetting is suppressed by the application of Al$_2$O$_3$ capping layers. Again, the area of dewetting was calculated and the results are shown in Figure 11b. These data match trends in Figure 10 with the 0.5 nm thick Al$_2$O$_3$ capping layer reducing the dewet area by 85%. Even one ALD pulse of AlO$_x$ is surprisingly efficient at suppressing dewetting.

In summary, the response of each sample during dewetting depends on the material choice of capping layer, its thickness and the power absorbed by the sample. For thicker capping layers of AlO$_x$ and Ta, the capping layer can survive the dewetting process—the Au simply dewets underneath the intact cap. Where the capping layer is thinner, surviving caps are not observed. In these cases, the Au must first break and then pull apart the capping layer. As the thinner capping layers provide optimal resistance against dewetting, it plausibly must be energetically more costly for the Au layer to break and pull the capping layer apart than it is to dewet underneath it. This supposition is supported by the model described in the following section. Notwithstanding, while capping layers of 0.5−1 nm proved to be superior in the present measurements, the addition of just one ALD pulse of Al$_2$O$_3$ reduced dewetting significantly.

**Energetics-Based Model for Capping Layer Protection.** To help understand the observed behavior on qualitative grounds, we next develop a simple model for the energetics of laser-induced dewet area opening in the laser absorbing layers. Such layers rest on a generic substrate below, with the option of a protective capping layer above. We will consider only because of its particularly thoroughly characterized parameters,[3] we will consider the Au(111)/Al$_2$O$_3$(1000) interface, with surface and interface energies as shown in Table 3. We use Al$_2$O$_3$ as both the adhesion layer and capping layer for our numerical example, although in the present experiments, SiO$_2$ was used. We will further consider a pore radius of $r$ = 1 μm (as seen in the experiment for more stable pores), the bulk Youngs modulus $Y$ = 380 GPa from Table 1 for Al$_2$O$_3$, a capping layer thickness of $t$ = 1 nm, together with a hillock radius of $L$ = 5 μm (that the distance from the pore center out to the somewhat radially distributed features seen, e.g., in Figure 5).

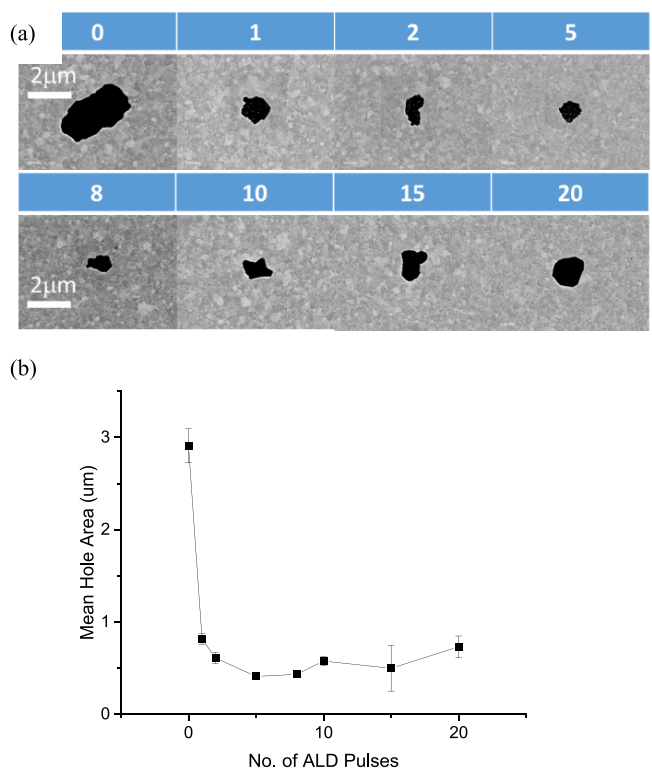

Figure 11. (a) SEM images of representative dewet areas of Ti 1 nm/Au 42 nm/Al$_2$O$_3$ capping layers, where the number of ALD pulses is indicated, post laser dewetting at $P_{abs}$ = 30 mW. (b) Plot of dewet areas as a function of number of Al$_2$O$_3$ monolayers $N$. Just one area was found in SEM for the $N$ = 5 sample. Error bars represent standard error.

Table 3. Surface and Interface Energies Used to Calculate Same Values from the Model[33]

| material | surface energies (J/m$^2$) | interface energies with Au (J/m$^2$) |
| --- | --- | --- |
| Au | 1.4 | |
| Al$_2$O$_3$ capping/adhesion | 1.24 | 2.15 ± 0.06 |

**Case A: Pore Creation in a Metal Layer Deposited on a Substrate.** We begin by considering the simple case A, depicted in Figure 12a. The system is an axially symmetric dewet area (a pore/hole, disk, or puck of vacuum or air) in an otherwise infinite sheet of metal denoted by M, which is deposited on a semi-infinite layer of substrate material denoted S. The circular surface of the exposed substrate has radius $r$ and area $A$, and the thickness, or height of the metal is $h$. The idealized energy differences between initial and final states, completely neglecting considerations of energy barriers and kinetics, and hence rates of change with time. The energy differences considered may be thought of as being averaged over many runs of an experiment or, equivalently by the ergodic principle, over many dewet areas observed in a large sample.

We will find that, despite its simplicity, this simple energetics-based model can explain our experimental observation that, while smaller metallic holes tend to open together with their capping layer, beyond a certain capping layer thickness, the capping layer tends to be left behind intact as a cap suspended over the hollow. The model further helps to explain why these intact capping layers do little to significantly arrest the further growth of dewet areas, as compared to the situation where no capping layer at all is provided. The most interesting result of the model, perhaps, is that the critical capping layer thickness at which capping layer preservation becomes energetically favorable scales inversely with the dewet area. Finally, the model will provide some nontrivial but easily implementable insights on how the substrate and overlayer materials might be best matched to a given strongly light-absorbing metallic layer that has high resistance to dewetting.

As we develop the model, it will prove helpful along the way to consider specific material and thin-film parameters, both as a check on signs and orders of magnitude, and to make connections to the experimental setup and observations. We reiterate here that the model is not a description of the dynamics or kinetics in this system, and we should typically not expect quantitative agreement. Indeed, for simplicity and surface energy of the metal is denoted by $\sigma_M > 0$, and that of the substrate is denoted by $\sigma_S > 0$. The metal–substrate interface energy is denoted by $\gamma_{MS}$, which may depend on the bonded crystal facet, but with lower values indicating a better match. The work (per unit area) of separation is then given by the standard expression $W_{MS} = \sigma_M + \sigma_S - \gamma_{MS}$.[32] Considering only surface effects, the energy cost of creating the situation depicted in case A from an infinite metal sheet with no dewet area is then given by

$$\Delta E^{(A)} = (\sigma_S - \sigma_M - \gamma_{MS})\pi r^2 + \sigma_M 2\pi r h \qquad (1)$$

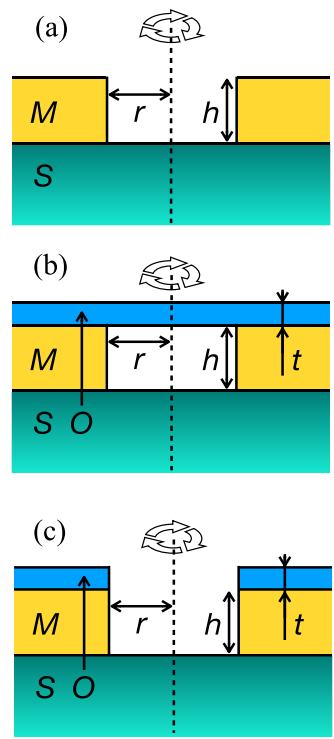

Figure 12. Schematic illustration of the three different cylindrical dewet area geometries considered in our energetics model. The materials are the metal M, that of the substrate, S, and that of the capping layer or over-layer, O. The geometry is symmetric about the axis of rotation shown by the dashed line. The key dewet area radius and thicknesses are shown. (a) Au film dewetting, (b), Au film dewetting leaving capping layer intact, and (c) Au dewetting and breaking the capping layer.

The expression $\Delta E^{(A)}$ represents the fact that, upon creating the dewet area, some substrate-vacuum surface area is exposed ($\sigma_S$ term), some metal-vacuum surface is removed ($-\sigma_M$ term), some metal–substrate interface area is removed ($-\gamma_{MS}$ term), and a strip of metal-vacuum surface is exposed around the circumference of the pore ($2\pi r$ term). For certain materials combinations, $\Delta E^{(A)}$ is always negative, suggesting that wetting would occur spontaneously at finite temperature, were not for kinetic barriers. Indeed, taking our numerical example and neglecting the annular term proportional to $h$, we find an endothermic $\Delta E^{(A)}$ of $-7 \times 10^{-12}$ J. To avoid this regime, we would need to have for all $r$ that

$$\Delta E^{(A)} > 0 \quad \Rightarrow \quad \sigma_S - \gamma_{MS} > \sigma_M\left(1 - \frac{2h}{r}\right) \qquad (2)$$

This condition requires $\sigma_S - \sigma_M > \gamma_{MS}$, a case which might be encountered in practice, but not for all materials combinations. If we can assume metal surface area conservation through hillock formation elsewhere, this simplifies to $\sigma_S > \gamma_{MS}$, which still may not be reliably assumed. This condition nonetheless provides a first rule-of-thumb for selecting viable substrates. While the surface energies of elemental metals have been comprehensively characterized,[33] their more challenging oxides have been much less so. As we do not observe spontaneous dewetting over long sample lifetimes and several samples, we may assume that kinetic barriers prevent that in reality. We do not include friction effects in our energy cost, the dominant contributions of which are expected to arise in the displacement outward of the metal layer, and in shear-like effects at the metal–substrate interface during that displacement. As these effects are common to the case A described and the two cases that will follow, they are not expected to significantly affect the relative energetic favorability of these cases, and the qualitative findings that are derived from the following comparisons.

Next, we suppose that the necessary net energy $\Delta E^{(A)} = E_L$ is provided by an incident laser. Here, $E_L$ is the net energy deposited by the laser during the period of pore formation, after reflection, heat dissipation and metal M diffusion, transport, and ultimate deposition are accounted for. In principle, while $E_L$ is ordinarily expected to be positive, it could even be negative for energetically favorable (endothermic) pore formation that is kinetically inhibited without the stimulus provided by the laser, without affecting the qualitative conclusions of the analysis that follows. Treating $E_L$ as a free parameter independent of the pore radius, in essence, we assume that the pore remains sufficiently comparable to the laser spot size that its growth does not appreciably affect the net power absorbed, which is of course an approximation. Eq 1 thereby yields a simple quadratic equation in the radius $r$, and we may further simplify that by neglecting the annular term proportional to metal height $h$. Then, defining the pore resistance constant $c_{MS} = \sigma_S - \sigma_M - \gamma_{MS}$, we obtain for the expected dewet area $A = \pi r^2$ the expression, in this case A,

$$A^{(A)} = \frac{E_L}{c_{MS}} \qquad (3)$$

Here, when $c_{MS}$ is negative then so necessarily will be $E_L$ (the endothermic case). Clearly, to minimize this area, for a noncapped metal surface M, it is best to maximize the surface energy $\sigma_S$ of the substrate, while minimizing the metal–substrate interface energy.

**Case B: A Capping Layer that Remains in Place as a Cap When a Dewet Area Is Formed.** Next, we will analyze case B, shown in Figure 12b, in which there is capping layer of material O deposited on top of the metal layer, and where this capping layer remains intact as a cap when the metal is transported away under laser irradiation. Given the third material, we introduce the surface energy $\sigma_O > 0$ and interface energies $\gamma_{MO}$ and $\gamma_{SO}$, and corresponding works of separation $W_{MO}$ and $W_{SO}$, with their obvious meanings. In the special case where O and S are the same material, then $\gamma_{SO} = 0$. Were the capping layer to fall into contact with the substrate, then an energy cost reduction of approximately $W_{SO}A$ would be gained, but this gain must be more than compensated by the cost of deforming or shearing the capping layer, since this geometry is not observed in our experiments. We therefore only treat the situation of capping layer over-suspension rather than the perhaps more obvious situation where the metal layer falls into contact with the substrate.

The energy cost of creating a pore in case B, with respect to leaving the metal and capping layer pristine, is given by

$$\Delta E^{(B)} = (\sigma_S + \sigma_O - \gamma_{MS} - \gamma_{MO})\pi r^2 + \sigma_M 2\pi r h \qquad (4)$$

This energy cost may again be negative in practice (evaluating to $-6 \times 10^{-12}$ J in our example, an insignificant change); however, a pore opening is generally less favorable with a capping layer present since $\Delta E^{(B)} - \Delta E^{(A)} = W_{MO}\pi r^2 > 0$. The corresponding expression for the expected dewet area, the cross-sectional area provided with a given laser-provided energy $\Delta E^{(B)} = E_L$ is, after neglecting the annular term and with some substitutions, given by the growth with $E_L$ (not to imply with power or time, recalling that kinetic barriers are neglected)

$$A^{(B)} = \frac{E_L}{c_{MS} + W_{MO}} \qquad (5)$$

This is not necessarily an improvement provided by the capping layer (leaving aside considerations of lateral shear, thermal insulation, etc.) since, for a given fixed $E_L$, we have that

$$\frac{A^{(B)}}{A^{(A)}} = \frac{c_{MS}}{c_{MS} + W_{MO}} \qquad (6)$$

which may be less than or greater than one, indicating some dewet area growth suppression or enhancement by the capping layer. For the material values given in Table 3, this evaluates to approximately 1.3, a marginal disimprovement in pore formation and metal damage suppression, but hardly a dramatic one. When making comparisons of this kind, we assume that the average energy to create a pore, over many pores in a sample, is independent of the presence of a capping layer. This energy is expected to be only a very tiny fraction of the total laser power; hence, the area expressions are only indicative and for relative comparison. The assumed independence of this energy with respect to pore size reflects the comparable spot size of the incident laser. While the presence of a capping layer does affect the reflectivity somewhat, this will only provide a multiplicative constant of order one in the previous equation, and not affect the qualitative outcomes of the analysis as we proceed. Clearly for a given metal and substrate, the task of choosing an overlayer in case B is that of maximizing the work of separation $W_{MO}$ between the metal and overlayer. In the case that the substrate surface energy is matched to that of the metal, so that $\sigma_S = \sigma_M$, and if we further assume that the same substrate material is used for the capping layer, then we have a modification, by adding that capping layer, of

$$\frac{A^{(B)}}{A^{(A)}} = \frac{-\gamma_{MS}}{W_{MS} - \gamma_{MS}} \quad (7)$$

which still evaluates to 1.3 in our example. Importantly, the linear character of the pore cross-sectional area A dependence on the deposited power is not affected by the introduction of a case B capping layer, and in general, surface energy considerations alone seem insufficient to explain how capping layers protect against pore formation.

**Case C: A Capping Layer that Resists Pore Formation Elastically.** The final geometry that we will consider is case C, depicted in Figure 12c, where the capping layer is pulled back with the metal to form part of the pore, with the same radius $r$. Given that typical capping layers will generally not melt, flow, or vaporize with the metal M but instead resist deformation, it is reasonable to add an energy cost for the dewet area opening in O. Perhaps the simplest cost term to assume is a linear one proportional to Y, an effective in-plane deformation modulus of the capping layer material O. We may assume that the radial compression in O takes place over a characteristic radial length L, beyond which the stress is relieved, e.g., by such surface eruptions. Hints of the physical value of L are the crater or halo radii that can clearly be seen in Figure 5, which importantly do not grow with dewet area based on our time-dependent observations. While we can think of the effective modulus Y as being akin to a Young's modulus, which would normally be lower for a film than for bulk, yet the effective modulus in question may be expected to be much greater in practice here. This is because Y must account for the possibly considerable compressive and shear stresses associated with the eruption and coverage of clearly evident hillocks of rejected material around the crater boundaries. The fact that the volume of such features will scale with film thickness $t$ means that their effect can be approximately absorbed into the effective modulus Y. In any case, we have (integrating strain over the dewet area radius, with increasing circumference as we go) that

$$E_{\text{elastic}} = Yt \int_0^r \left(\frac{s}{L}\right)^2 (2\pi s) ds = \frac{Yt}{L^2} \frac{\pi r^4}{2} \quad (8)$$

This expression is derived by considering the work done against an inward stress upon the wall of a cylindrical pore region of thickness $t$ and variable radius $s$, which creates a radial in-plane effective strain $s/L$, where L is the radius over which the compression (and related forces included in the effective modulus Y and related, e.g., to perimeter hillock formation) occurs. A further factor of $s/L$ arises to account for the ratio of arc length for a given segment at the expanding radius $s$, and the fixed radius $L$, that is due to the fact that smaller volumes are being compressed at a lower pore radius. Integrating the radius up to its maximum value of $r$, while noting that the cylinder curved surface area expands with the circumference, yields the final result. An alternative way to arrive at the same result is perhaps to consider the standard quadratic energy of compression $E_{\text{elastic}} = \left(\frac{Y}{2}\right)\frac{(V-V_0)^2}{V_0}$, which gives the same result as eq 8 when we consider the total volume $V_0 = t\pi L^2$ being compressed to $V = t\pi L^2 - t\pi r^2$ upon the opening of the pore. Clearly, since for example, $E_{\text{elastic}}$ vanishes for infinite L (vanishing strain) and this may be considered pathological. It may be possible to further elaborate on the elastic forces at play; however, the present model serves as a first approximation here. The total energy cost of creating a pore in case C is then given by

$$\Delta E^{(C)} = (\sigma_S - \sigma_O - \gamma_{MS} - \gamma_{MO})\pi r^2 + (\sigma_M h + \sigma_O t) 2\pi r + \frac{Yt}{L^2} \frac{\pi r^4}{2} \quad (9)$$

This energy cost will certainly be positive beyond a sufficient radius $r$. Setting $\Delta E^{(C)} = E_L$ results in a quartic equation for $r$, which may be solved in principle, but also approximated. For all but the thickest overlayers and narrowest dewet areas, the term proportional to $\sigma_0 t$ may be neglected. In this case, and taking our numerical example, the elastic term eq 8 evaluates to $+2.4 \times 10^{-11}$ J. This dominates in eq 9, which totals $+1 \times 10^{-11}$ J, within which the elastic term is twice as large as the opposing interface energy term. Even if the elastic term can be neglected, then case C is always energetically favored over case B, since then $\Delta E^{(B)} - \Delta E^{(C)} = 2\sigma_0 A > 0$. However, when Y is nonzero, case B is favored at sufficient $r$.

A simple check that can be made, next, is for the radius beyond which the elastic term dominates (recall that we are not envisaging here a temporal process, but only making before-and-after dewet area formation total-energy comparisons). For this, we neglect the remaining linear term and set the quadratic and quartic terms to be equal and opposite, yielding

$$r_{\text{elastic}}^2 = -\frac{2(\sigma_S - \sigma_O - \gamma_{MS} - \gamma_{MO})L^2}{Yt} =_{O=S} \frac{4\gamma_{MS}}{Y}\frac{L^2}{t} \quad (10)$$

Given the material values in Table 3 and taking our numerical example where O = S, this $r_{\text{elastic}}$ evaluates to 0.8 $\mu$m, which is comparable or perhaps one order of magnitude below the observed pore sizes. This supports the conclusion that pore growth appears to be arrested not far into the elastic regime, which is physically plausible. It also, again, reflects that it is advantageous to minimize $\gamma_{MS}$ if possible, to increase the energetic cost of pore formation. Now, of course, the film Youngs modulus will be lower than this bulk value, yet the effective modulus Y is expected to be much higher, increasing this energy. Nonetheless, the analysis demonstrates that cases B and C may compete in the experimental regime.

Since cases B and C may compete thermodynamically, let us next estimate the radius at which the energetic crossover from one regime to the other occurs. Setting $\Delta E^{(B)} = \Delta E^{(C)}$ and solving for $r$, we arrive at

$$r_{crossover}^2 = \frac{4\sigma_O}{Y}\frac{L^2}{t} \quad (11)$$

In our numerical example, $r_{crossover}$ evaluates to 0.6 μm. Which of $r_{crossover}$ and $r_{elastic}$ is greater is materials-specific, and not critical. The key insight rather, from inverting eq 11, is that that the critical capping layer thickness at which capping layer preservation becomes energetically favorable scales inversely with the dewet area. Indeed, by inverting to find the capping layer thickness $t$ up to which the more dewetting-resistant case C survives, we can then evaluate that using the simpler $E_L$ dependence on $r$ in the case B regime just beyond, from eq 5, to give

$$t = \frac{4\sigma_O}{Y}\frac{\pi L^2}{\pi r_{crossover}^2} = \frac{4\sigma_O}{Y}\pi L^2 \frac{c_{MS} + W_{MO}}{E_L} \quad (12)$$

This means that the critical thickness shrinks with increasing absorbed energy, if $L$ can be considered constant, and also thereby scales inversely with the dewet area. As a more approximate, simpler approach, we might invert eq 11 while setting $r_{crossover}$ to half of the pore size, 0.5 μm say, to evaluate a capping layer thickness below which dewetting through case B should be disfavored, 1.3 nm. This concurs with the experimentally observed orders of magnitude, which is the most that we can be expected in terms of quantitative agreement, recalling the simplicity of the model with respect to experimental reality.

**Overall Picture Emerging from the Simple Energetics Model.** We can overall envisage three regimes, to summarize, with increasing dewet area radius $r$ (not increasing in time, but when samples are averaged). First, case C simply dominates, before the capping layer elastic energy term starts to overwhelm the other costs of creating case C dewet areas. In some material cases, and perhaps in some regions depending on the local defect and grain structure, case B (with the overlayer being left behind as a cap) may become energetically favored, reflecting experimental observations. That it is necessary to maximize the capping layer surface energy $\sigma_O$, to maximize the dewet area sizes over which case C is maintained, is unsurprising. What is surprising, however, is that it is best for that purpose to minimize the layer's in-plane effective modulus $Y$ and/or its thickness $t$, so that case C is sustained up to larger radii. It is unclear what factors influence the stress relief radius $L$; however, the laser spot size, grain structure, and thermal factors may play a role.

Supposing that we remain in case C, we may again compare the dewet area cross-sectional area A for a given input power $E_L$, against that with no capping layer, case A. Supposing first small energy and small dewet areas so that the elastic regime is not reached in case C, we have

$$\frac{A^{(C)}}{A^{(A)}} = \frac{c_{MS}}{c_{MS} + W_{MO} - 2\sigma_0} \quad (13)$$

For the materials parameters in Table 3, this comes to 0.54, representing an almost factor-of-two improvement over case B, which is hardly significant. Again, this is leaving aside elasticity. If we again assume, for simplicity, that the substrate surface energy is matched so that $\sigma_S = \sigma_M$, and furthermore that O = S, then

$$\frac{A^{(C)}}{A^{(A)}} = \frac{-\gamma_{MS}}{W_{MS} - \gamma_{MS} - 2\sigma_S} \quad (14)$$

For the same material values, this again evaluates to 0.5, rendering the capping layer essentially useless in this small-area limit where the elastic term is not yet relevant. It is clear then, given that capping layers are advantageous in practice, that an elastic term or a similar deformation penalty for the capping layer plays a key role in limiting dewet area growth.

Let us suppose next that we are in the elastic regime then, beyond $r_{elastic}$. Setting $E_L = \Delta E^{(C)}$, we find that the dewet area in case C grows with $E_L$, yet shrinks with $Y$ and $t$, like

$$A^{(C)} = \left(\frac{2\pi E_L L^2}{Yt}\right)^{1/2} \quad (15)$$

and, thereby, that the pore area reduction provided by a stretchable capping layer is strongest for largest $Y$, $t$, and absorbed laser energy $E_L$. Here, we note that such power-law like behavior is certainly not evident throughout Figure 6, but perhaps is visible in the expected subnanometer regime, however where it could also be attributable to partial capping layer coverage. If we look at the most promising data points for $AlO_x$, in Figure 6b, for the highest wattage, we can estimate a power absorbed, we do indeed see a close to $t^{-0.5}$ behavior, however inconclusively. Using eq 15 and our example parameters, from this data, we arrive at a cost of pore formation of approximately $+1.5 \times 10^{-11}$ J (= $+9 \times 10^7$ eV in perhaps more nanoscale-relevant units), coming down to $+1.0 \times 10^{-11}$ J when the surface term of eq 9 is included.

There is, in the end, a trade-off at play. If case B were not available, one would simply maximize the capping-layer thickness $t$ to minimize the pore area. However, doing this makes case B relatively more likely, and this results in ultimately larger pore growth. When the model is combined with the experimental observations, we are led to the conclusion that the advantages of larger capping layer thicknesses $t$ are more than compensated for by the risks of separation between metal and capping layer. The largest dewet area radius that avoids break down to case B is $r_{crossover}$, and this corresponds to an absorbed energy, if we can neglect the linear terms within the case c regime, of

$$E^{(C)}(r_{crossover}) = (\sigma_S + \sigma_O - \gamma_{MS} - \gamma_{MO})\sigma_O \frac{4\pi L^2}{Yt} \quad (16)$$

Thus, the energy that can be absorbed while maintaining capping layer control, for case C pores, scales quadratically with $\sigma_O$ and with the inverse of both $Y$ and $t$. This expression agrees with the limit from case B, and this again means that the critical thickness shrinks with increasing absorbed energy, if $L$ can be considered constant. In our numerical example, eq 16 evaluates to a modest $-2 \times 10^{-12}$ J.

It is worth finally considering again the $r_{crossover}$ between cases B and C (eq 11). In any real system, there is no benefit in maintaining the cap while the Au underneath dewets. We find a value as high as 0.6 μm = 0.6 $r$ in this example, again assuming the Young's modulus value for $Y$, but in a thin layer of 1 nm. This reflects that the risk of case B rupture is not remote, indeed as we observe experimentally. The conclusion is that, even a very in-plane stretchy, thin capping layer is better than an in-plane stiff one, if it means that the capping layer can remain securely in contact with the metal. Somewhat surprisingly, this suggests the use of a capping layer material with strong out-of-plane bonds (internally, and with the metal), even if with relatively weak or deformable in-plane bonds if that compromise is necessary. However, if the emergence of case B can be fully suppressed by other means or mechanisms (e.g., a further overlayer, or shear resistance), then conversely high $t$ and $Y$ would of course be preferred to maximize the penalty of overlayer elastic deformation.

Furthermore, our analysis shows that increasing the distance $L$ beyond which strain may be released, e.g., via surface welling-up or out-flow, is equivalent to minimizing the capping layer thickness $t$, for the purposes of suppressing case B. Indeed, this is in qualitative agreement with the finding of Cao et al.[23] that monolayer graphene can help arrest Au dewetting, since it is very thin and offers very good coverage (if case B is assured, then its very high effective Young's modulus is an asset).

To summarize, our analysis leads us to propose several strategies for together optimizing both the substrate material S and the capping layer material O, for a given metal M of thickness $h$:

- Minimize the interface energy $\gamma_{MS}$ and maximize the substrate surface energy $\sigma_S$, which together amount to maximizing the work of separation $W_{MS}$.
- To preserve case C up to the largest dewet areas, maximize the capping layer surface energy $\sigma_O$ and minimize the in-plane Young's modulus (or effective deformation modulus $Y$, if it can be estimated). Minimize the capping layer thickness $t$, short of compromising the continuity of that film. Indeed, maximize the continuity (grain size) of the overlayer if possible, if we can suppose that may also increase the elastic length scale $L$.
- If suppression of case B in favor of case C could be assured by other means, we conjecture that it would be best to conversely maximize both the in-plane Young's modulus (or effective deformation modulus $Y$, as may be appropriate) and capping layer thickness $t$.
- As a by-far secondary consideration with respect to maximizing $\sigma_O$, yet a valid consideration in both cases B and C is to maximize $W_{MO}$, i.e., minimize $\gamma_{MO}$.

## CONCLUSIONS

In this work, the dewetting response of adhesion layer/gold films with a variety of capping layers was examined. For optimal dewetting resistance, thinner capping layers offer greatest protection and, under certain circumstances, dewetting can be entirely prevented. Detailed SEM and AFM investigations of the dewet areas reveal that the Au film can dewet underneath thicker, continuous, elastic capping layers, leaving behind a suspended membrane of material. In the case of thinner capping layers (1 nm and below), there likely exists a discontinuous island morphology. This enables these capping layers to dewet as the Au film dewets. As thinner capping layers are shown to result in least damage to the Au film, it is apparent that pulling these discontinuous capping layers apart is energetically more costly than simply dewetting underneath thicker, continuous variants, and that thinner capping layers are optimal because their combination of thickness, adhesion, and mechanical strength force the retreating Au film to pull the capping layer with it as it dewets.

A simple model based on energy differences, neglecting kinetic effects, has allowed us to explain the key experimental finding that thinner capping layers, surprisingly, tend to offer better protection to metal layers against the formation of laser-induced dewet areas, and thereby the subsequent more extensive damage. The energetic competition between the two most commonly observed area geometries is a key element of the model, namely, case B where the capping layer is left behind suspended over the metal pore and case C where the capping layer pulls away with the dewet metal. We explain how case C is better able to absorb input energy by storing it elastically, and why a thinner capping layer allows this more protective geometry to energetically prevail up to larger dewet area radii. In fact, we predict that the energy that can be absorbed, before the dewet area radius is reached where the capping layers tend to fail, grows with the inverse of the capping layer thickness and of its in-plane elastic modulus. We furthermore identify the capping layer surface energy as being a key quantity to maximize when choosing such materials. We provide a hierarchical list of model-derived rules-of-thumb, given a metal to be protected, for optimizing the choice of substrate and capping layer materials. This model is amenable to further extension and refinement in future, e.g., by considering irregularly shaped dewet areas and their statistical averages, variation in reflectivity, power absorption, and heat dissipation with dewet area geometry, and more elaborate treatments of elasticity and/or plastic deformation.

In conclusion, this work demonstrates and explains the positive impact on application of thin (<5 nm) capping and adhesion layers on Au plasmonic films and offers a path toward increased reliability in applications such as HAMR.

## METHODS

**Sample Preparation.** Quartz substrates were cleaned by means of sonication in acetone, then isopropanol, then dried using an $N_2$ pistol. This was followed by a 3 min plasma clean in an oxygen plasma asher before deposition. Samples were deposited using a 6-source SHAMROCK 19608DC/RF sputtering system, which has a base pressure of $5 \times 10^{-7}$ Torr. Deposition rates for Ti, Ta, Au, $Al_2O_3$, and Al were first calculated by measuring the thickness of timed depositions using low-angle X-ray reflection (Phillips X'pert Pro XRD with Cu $k\alpha$ radiation). Reliable deposition rates range from 0.08 to 0.36 Å s$^{-1}$. Using this system, films were deposited sequentially at ambient temperature in mTorr Ar partial pressures without breaking vacuum. As actual film thicknesses for adhesion and capping layers cannot be measured directly, they should be regarded as target thicknesses.

Some samples were capped using atomic layer deposition employing a Picosun R200 ALD reactor, using trimethylaluminum (TMA) as the Al precursor and water as the coreagent in a metal pulse first process. Precursor pulses for both reagents were set at 0.1 s with 10 s purges at a reactor temperature of 150 °C.

**Sample Characterization.** Absorption spectra for the films were obtained using a Perkin Elmer UV–vis spectrophotometer. For each film, the reflectivity and transmission were measured, which then allowed the absorption to be calculated from the relation $A = 100 - R - T$. This ensured that the absorbed power was consistent across all samples.

*In situ* measurement of dewetting dynamics of sputtered thin films we performed using an optical technique described in detail previously.[11] The use of a focused COHERENT INNOVA 90C Ar + laser operating at 488 nm was directed into the back of an Olympus Plan 0.4NA microscope objective to focus the laser ($\omega_0 = 1.8 \pm 0.1$ μm) onto the sample as a localized, micron-scale heat source. This closely represents the expected heat sources and resultant thermal gradients in applications such as HAMR. The focused laser spot is used to induce solid-state dewetting causing local reduction in the film reflectivity. This resultant reduction in the back-reflected laser signal is measured over time to show the progress of solid-state dewetting.

SEM images were obtained using a Zeiss ULTRA scanning electron microscope equipped with a GEMINI FESEM column capable of 1 nm resolution at 15 kV, using the SE2 detector. The beam voltage was 5 kV for all images. AFM measurements were performed using an Asylum MFP-3D. AFM was used in tapping mode with Budget Sensors probe type Tap300Al-G. Scan rates were 0.5–1 Hz.

## ASSOCIATED CONTENT

### Supporting Information

The Supporting Information is available free of charge at https://pubs.acs.org/doi/10.1021/acsaelm.3c00052.

    Additional AFM images of dewet samples (PDF)

    Video of a thin film sample dewetting under laser radiation at 3× normal speed (MP4)

## AUTHOR INFORMATION

### Corresponding Author

    **Christopher P. Murray** − *School of Physics, CRANN and AMBER, Trinity College Dublin, The University of Dublin, Dublin 2, Ireland*; orcid.org/0000-0002-7496-9442; Email: murrayc2@tcd.ie

### Authors

    **Daniyar Mamyraimov** − *School of Physics, CRANN and AMBER, Trinity College Dublin, The University of Dublin, Dublin 2, Ireland*

    **Mugahid Ali** − *School of Physics, CRANN and AMBER, Trinity College Dublin, The University of Dublin, Dublin 2, Ireland*

    **Clive Downing** − *School of Physics, CRANN and AMBER, Trinity College Dublin, The University of Dublin, Dublin 2, Ireland*

    **Ian M. Povey** − *Tyndall National Institute, Cork T12 R5CP, Ireland*; orcid.org/0000-0002-7877-6664

    **David McCloskey** − *School of Physics, CRANN and AMBER, Trinity College Dublin, The University of Dublin, Dublin 2, Ireland*; orcid.org/0000-0003-3227-5140

    **David D. O'Regan** − *School of Physics, CRANN and AMBER, Trinity College Dublin, The University of Dublin, Dublin 2, Ireland*; orcid.org/0000-0002-7802-0322

    **John F. Donegan** − *School of Physics, CRANN and AMBER, Trinity College Dublin, The University of Dublin, Dublin 2, Ireland*; orcid.org/0000-0002-5240-1434

Complete contact information is available at:
https://pubs.acs.org/10.1021/acsaelm.3c00052

### Author Contributions

The manuscript was written through contributions of all authors. All authors have given approval to the final version of the manuscript.

### Notes

The authors declare no competing financial interest.

## ACKNOWLEDGMENTS

This work was funded by Science Foundation Ireland (SFI) grant no. SFI/12/RC/2278_R2 (WMA, CPM, CZ), and ASRC. This work was further supported by Science Foundation Ireland [19/EPSRC/3605] and the Engineering and Physical Sciences Research Council [EP/S030263/1] (DDOR). The authors would like to thank G. Atcheson for assistance with film deposition. We are grateful to John Boland and Jon Sader for discussions on dewetting processes.

## ABBREVIATIONS

AFM, atomic force microscopy; EDX, energy-dispersive X-rays; HAMR, heat-assisted magnetic recording; SEM, scanning electron microscopy; XRR, X-ray reflectometry